\documentclass{mem}
\usepackage{natbib}\usepackage{txfonts}\usepackage{balance}
\usepackage{graphicx}
\usepackage{txfonts}
\usepackage[a4paper]{hyperref}
\idline{79}{3}
\begin{document}
\def\teff{$T\rm_{eff }$}
\def\leoii{\hbox{Leo\,II}}
\def\hi{\ion{H}{i}}
\def\hii{\ion{H}{ii}}
\def\caii{\ion{Ca}{ii}}
\def\msol{${\mathcal M}_{\odot}$}
\def\lsol{${\mathcal L}_{\odot}$}
\def\kms{km s$^{\rm -1}$}
\def\muchless{\ll}

\title{
The   blue plume population in  dwarf spheroidal galaxies: 
genuine blue stragglers or young stellar population?}

   \subtitle{}

\author{
Momany \, Y.\inst{1} 
\and
Held, E.V.\inst{1}
\and
Saviane, I.\inst{2}
\and
Zaggia, S.\inst{1}
\and
Rizzi, L.\inst{3}
\and
Gullieuszik, M.\inst{1}
}

  \offprints{Y. Momany}

\institute{
INAF: Oss. Astronomico di Padova, 
vicolo dell'Osservatorio 5, 35122 Padova, Italy 
\and
European Southern Observatory, A. de Cordova 3107, 
Santiago, Chile
\and
Institute for Astronomy, 2680 Woodlawn Drive, Honolulu, 
HI 96822, USA
\email{yazan.almomany@oapd.inaf.it}
}

\authorrunning{Momany}

\titlerunning{BSS in dwarf galaxies}

\abstract{
In the  context  of dwarf spheroidal galaxies   it is hard   to firmly
disentangle a  genuine Blue Stragglers  (BSS) population from a normal
young main (MS) sequence. This difficulty is persistent.
For a sample   of 9 non-star  forming  Local Group  dwarf  galaxies we
compute the ``BSS frequency'' ($F^{\rm BSS}_{\rm HB}$) and compare it with
that found in the Milky Way globular/open clusters and halo.
The comparison shows that $F^{\rm BSS}_{\rm HB}$ in dwarf galaxies, at
any given ${ M_{V}}$, is always  higher than that in globular clusters
of similar  luminosities.  Moreover, the  estimated $F^{\rm  BSS}_{\rm
HB}$ for  the   lowest  luminosity  dwarf   galaxies  is  in excellent
agreement with that observed in the Milky Way halo and open clusters.
We conclude that  the low density, almost  collision-less environment,
of our dwarf galaxy sample point to their very low dynamical evolution
and consequent negligible production of collisional BSS.
}
\maketitle{}

\section{Introduction}

First identified by Sandage   \cite{sandage53}, Blue Stragglers  are
usually defined  as  a  hotter and  bluer   extension  of  normal main
sequence  stars)
The origin of BSS is sought as either {\it primordial binaries} coeval
with  the globular/open  cluster formation epoch,   or to a continuous
production (in successive epochs) of {\it collisional binaries} due to
dynamical collisions/encounters  experienced    by single/binary stars
throughout the life of the cluster.

Piotto et  al.  \cite{piotto04} presented a  homogeneous compilation
of $\sim3000$ BSS (based on HST  observations of 56 globular cluster),
and derived a significant and rather puzzling anti-correlation between
the  BSS specific frequency and  the cluster total absolute luminosity
(mass).
That  is  to say   that more  massive  clusters are  surprisingly  BSS
deficient, as if their higher collision rate had no correlation with
the production of collisional BSS.
Another puzzling observable  is  that the BSS  frequency in  Milky Way
field  (Preston \&  Sneden  2000) is  at  least an order of  magnitude
larger than that of globular clusters.

In the context of  dwarf galaxies one  {\em cannot} exclude  that blue
plume stars may include genuinely young main sequence (MS) stars, i.e.
a residual star forming    activity (e.g.  Held 2005,  and  references
therein). 
A useful indicator of  the presence of a   recent star formation in  a
galaxy  is the detection of a  vertical extension in correspondence of
the red HB region.   Stars  forming this  sequence are usually  called
vertical clump stars (VC, Gallart et al. 2005).
These are helium-burning stars  of few hundred  Myr to $\sim1$ Gyr old
population whose progenitors are to be searched in the blue plume.
Nevertheless, we note that Ferraro et al.  \cite{ferraro99} identify a
similar sequence in  the M80 globular cluster   and ascribe it to  the
evolved-BSS population.  We  therefore  cautionly conclude  that   the
detetction of VC star in a dwarf galaxy is not a clear-cut evidence of
the presence of a young MS population.

To shed  light on the BSS-MS ambiguity  in  dwarf galaxies, we measure
the BSS frequency in 9 non-star  forming dwarf galaxies and compare it
with that derived in other stellar systems.
Dwarf spheroidals/irregulars in which there is {\em current} or recent
($\le500$ Myr) star formation are not considered.
Dwarf spheroidals/irregulars showing  evidence of  recent $\le500$ Myr
star formation (e.g.  Fornax  dwarf, Saviane et al. 2000),
or young  MS  stars reaching the  horizontal  branch [e.g. Sagittarius
irregular, Momany et al. \cite{momany05}] level have been excluded. 
The  Canis Major dwarf  galaxy (Martin et   al. 2004 and Bellazzini et
al.  2004)  was   not  included [see    arguments   in Momany et   al.
\cite{momany04w} and \cite{momany06w}].
The  Carina dwarf (Monelli  et al.    2003),  is one of those  showing
evidence  of star  formation in  recent epochs  ($\sim1$ Gyr),  and we
include in our  sample so as to compare  with other  dwarf galaxies of
similar luminosities.

\section{The dwarf galaxy sample}
The BSS frequency  was derived  for  Sagittarius, Sculptor,
Leo~II, Sextans, Ursa Minor, Draco, Carina, Ursa Major and Bo{\"o}tes.
The Sagiitarius   data are based  on $BVI$ $1^{\circ}\times~1^{\circ}$
WFI@2.2m    data,  from       which    we    excluded      the   inner
$14\farcm\times~14\farcm$ region  around M54.  These were  reduced and
calibrated following the standard recipes in Momany et al.
\cite{momany01eis} and \cite{momany02sag}.
For  the remaining dwarf galaxies we  estimate the  BSS frequency from
either public photometric  catalogs (Sextans by  Lee  et al. 2003)  or
photometry kindly  provided by the  authors (Ursa Minor  by Carrera et
al.   2002, Draco by Aparicio et  al. 2001, Sculptor  by  Rizzi et al.
2003, Ursa  Major by Willman et al.   2005, Bo{\"o}tes by Belokurov et
al.  2006 and Carina by Monelli et al.  2003).

All  the photometric catalogs  extend  to and  beyond the galaxy  half
light  radius; i.e.  we cover a  significant  fraction of the galaxies
and therefore the estimated  BSS frequency should  not be  affected by
specific spatial gradients, if present.    The only exception is  that
relative to Sagittarius.
With a  core radius of $\sim3.7^{\circ}$,  the estimated BSS frequency
of our $1^{\circ}$ square degree field refers to less than 3.5\% areal
coverage of Sagittarius, or   a conservative $\sim6$\% of  the stellar
populations. Therefore the Sagittarius  BSS frequency  should
be considered with caution.
In order to account  for the foreground/background contamination, star
counts  were also  performed  on simulated   diagrams (using  the {\sc
Trilegal} code Girardi et al.  2005) and these were subtracted from
the BSS and HB star counts for the dwarf galaxy sample.
We calculate the specific frequency of  BSS (normalizing the number of
BSS to the HB) as: $F_{\rm HB}^{\rm BSS}={\rm \log}(N_{\rm BSS}/N_{\rm
HB})$.
%

\begin{figure*} 
\centering
\includegraphics[width=12cm,height=11cm]{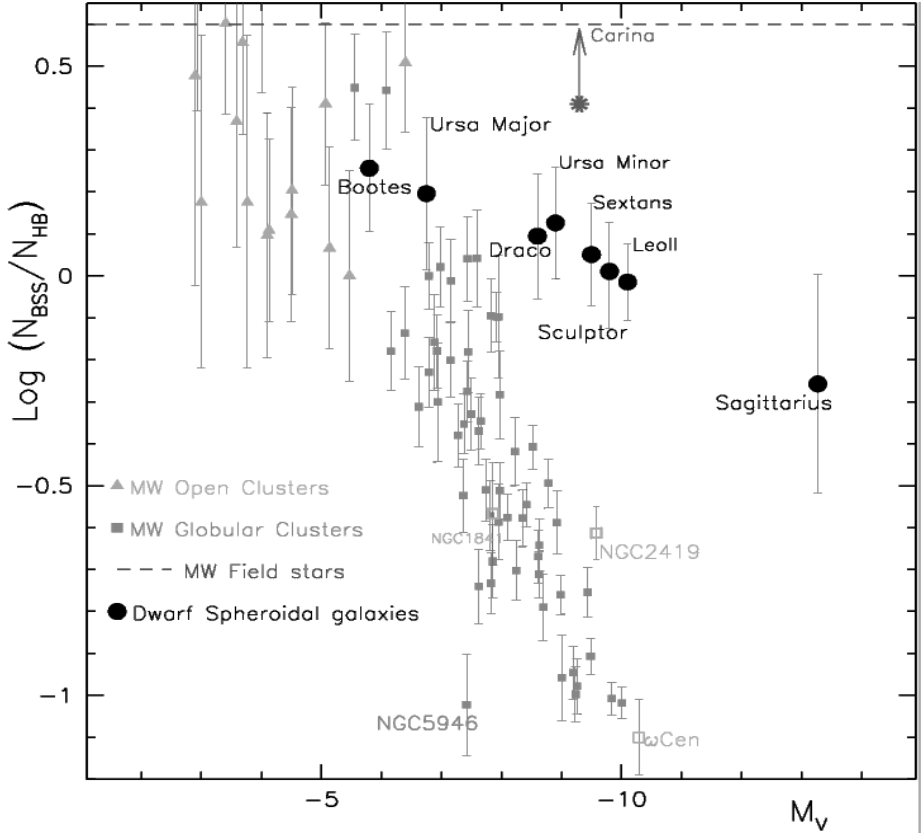}
\caption{The $F_{\rm BSS}$ vs ${ M_{V} }$ diagram for globular
clusters (Piotto et al.  2004) open clusters (De  Marchi et al.  2006)
and dwarf spheroidal galaxies.  The horizontal line shows the mean BSS
frequency for Milky Way field stars (Preston \& Sneden 2000). }
\label{fig1}
\end{figure*}

\section{BSS frequency in dwarf galaxies and globular clusters}

Figure~\ref{fig1}  displays the  $F^{\rm BSS}_{\rm HB}$ {\it  vs}   ${ M_{V} }$
diagram for our  dwarf galaxy sample  together with the data-points of
Piotto et al.  \cite{piotto04} and De Marchi et al.
\cite{demarchi06} for globular and open clusters, respectively.
Of the original  open cluster sample we only  plot clusters for  which
$\ge2$ BSS stars were found.
To the globular cluster sample we add the BSS frequency of $\omega$Cen
as  derived by Ferraro  et al. \cite{ferraro06},   and that of NGC1841
(Saviane et al.   2003) the LMC  most metal-poor and most  distant LMC
globular cluster.
Figure~\ref{fig1} clearly  shows  that, regardless  of their  specific
peculiarities, $\omega$Cen and NGC1841 are consistent with the general
globular clusters $F^{\rm  BSS}_{\rm HB}-{ M_{V}  }$ anti-correlation,
adding only universality to it.

As  for the dwarf galaxy sample,  it results immediately that the {\em
lowest luminosity dwarfs} (Bo{\"o}tes and Ursa  Major) would possess a
higher $F_{\rm HB}^{\rm BSS }$  than globular clusters with similar ${
M_{V} }$.
Most interestingly, their  $F_{\rm  HB}^{\rm BSS }$  is  in fact fully
compatible with that observed in open clusters.
This compatibility   between   dwarf galaxies and   open  clusters may
suggest  that there exists a  ``saturation'' in the  BSS frequency (at
$0.3-0.4$) for  the lowest luminosity systems.  
Thus, the relatively  high $F_{\rm HB}^{\rm  BSS }$  of Bo{\"o}tes and
Ursa   Major adds    more   evidence in   favor  of  a    dwarf galaxy
classification of the 2 systems.
Indeed, although their   luminosities is  several times fainter   than
Draco or  Ursa   Minor,  the  physical  size   of  the  two   galaxies
($r_{1/2}\simeq220$ and  $250$  pc respectively)  exceeds that of more
massive galaxies like Ursa Minor ($r_{1/2}\simeq150$ pc).

Another interesting feature is  the significant difference between the
BSS frequency  of Carina with  that  derived for dwarf  galaxies  with
similar luminosity, i.e.   Draco,  Ursa Minor, Sextans, Sculptor   and
Leo~II.
Although it is only a lower limit, the ``BSS'' frequency for Carina is
of great help   in  suggesting a threshold   near which  a  galaxy BSS
frequency might   hide   some level of   recent   star formation.  The
aforementioned 5  galaxies however have a  lower BSS frequency, a hint
that  these galaxies  possess a normal   BSS population rather  than a
young MS.

Lastly, leaving aside the extreme dynamical history of Sagittarius and
allowing    for   the  uncertainties  (due    to  the   heavy Galactic
contamination and the  relatively small sampled  populations) it turns
out  that its blue   plume-HB frequency is (i)  lower  than that in  a
star-forming galaxy like Carina,  and most interestingly; (ii) in good
agreement with the  expected BSS frequency  (note a hint of  a $F^{\rm
BSS}_{\rm HB}-{ M_{V}}$  anti-correlation for the 7 remaining galaxies
in our sample).
Added  to the clear absence of  MS stars overlapping  or exceeding the
Sagittarius HB luminosity level,  we suggest that the Sagittarius blue
plume is   a  ``normal'' BSS sequence.     As a matter  of  fact, {\em
Sagittarius  is   probably the nearest   system with  the  largest BSS
population:        over   2600      BSS     stars    in    the   inner
$1^{\circ}\times~1^{\circ}$ field}.

\section{ A $F^{\rm  BSS}_{\rm   HB}-{ M_{V}}$ anti-correlation for
dwarf galaxies ?}   

We here explore the  statistical  significance of a  possible  $F^{\rm
BSS}_{\rm   HB}-{\rm  M_{V}}$    correlation.   The linear-correlation
coefficient (Bevington 1969)   for the 8  galaxies  (excluding Carina)
data-points is $0.984$.
The corresponding probability that any  random sample of uncorrelated
experimental  data-points   would yield a correlation   coefficient of
$0.984$ is $<10^{-6}$.
Given the  greater uncertainties associated   with the Sagittarius BSS
frequency,  one  may  be  interested  in the   correlation coefficient
excluding the Sagittarius data-point.
In this case,   the resulting correlation coefficient  remains however
quite   high   ($0.972$) and  the probability     that the 7 remaining
data-points would randomly correlate is as low as $10^{-4}$.
Thus,  the  statistical significance of  the   $F^{\rm BSS}_{\rm HB}-{
M_{V}}$  anti-correlation in non  star-forming dwarf galaxies is quite
high.
We    follow  the   methods      outlined   in Feigelson   \&     Babu
(1992) and  fit least-squares  linear regressions.
In particular, the  intercept and slope regression coefficients  were
estimated through 5 linear models  (see the code  of Feigelson \& Babu
for         details)     the       average       of     which    gives
($a,b$)$=$($0.699\pm~0.081,0.070\pm~0.010$)                        and
($a,b$)$=$($0.631\pm~0.120,0.062\pm~0.014$)   including and  excluding
the Sagittarius data-point, respectively.
The  reported errors were  estimated  through {\sc Bootstrap} and {\sc
Jacknife} simulations so  as  to provide more   realistic $a$ and  $b$
errors.

However,   to firmly establish  this   $F^{\rm BSS}_{\rm HB}-{ M_{V}}$
anti-correlation one needs  to increase the  dwarf galaxies sample, in
particular at the two luminosity extremes.
Unfortunately there are not many non  star-forming dwarf galaxies with
$-13.3\le~M_{V}\le~-10.1$, and few  exceptions  may come from   deeper
imaging of galaxies like And~I and And~II.
On   the  other hand,   more  Local Group  dwarf   galaxies are  being
discovered in the low luminosity regime ($-8.0\le~M_{V}\le~-5.0$), and
it is necessary to include these in any analysis similar to ours.

\section{Conclusions}
For a sample of 8 non star-forming dwarf galaxies,  we have tested the
hypothesis that the blue plume  populations are made  of a genuine BSS
population (as   that observed in   open and   globular clusters)  and
estimated their frequency with respect to HB stars.
Should this assumption be incorrect (and  the blue plume population is
made of young MS stars) then one  would not expect an anti-correlation
between the   galaxies  total luminosity   (mass) and  the  blue plume
frequency, but rather a correlation between the two.
Instead, and within the limits of this and similar analysis, we detect
a statistically significant anti-correlation between $F^{\rm BSS}_{\rm
HB}$ and $M_{V}$.
Thus, should  a dwarf galaxy  ``obey''  the $F^{\rm BSS}_{\rm  HB}-{ M_{V}}$
anti-correlation   displayed by our     sample then  its blue    plume
population is probably made of blue stragglers.

We  also    estimated  stellar  specific   collision parameter  (${\rm
log}~\Gamma_{\star}$: the number of collisions per star per year).
The   mean collisional   parameter  of   the  9  studied   galaxies is
$\simeq-19$.  The  lowest value  is  that  for Sagittarius with  ${\rm
log}~\Gamma_{\star}\simeq-20.2$, and this is due  to its very extended
galaxy core.
Compared with  the  mean value  of $-14.8$  for the  globular clusters
sample   [see  lower   right  panel   of     Fig.~3  in  Momany     et
al. \cite{momany07}] the estimated number of  collisions per star per
year in a dwarf spheroidal is $10^{-5}$ times lower.

To summarize, from Fig.~\ref{fig1}  one finds that  $F^{\rm BSS}_{\rm
HB}$ in  dwarf galaxies  is (i)  always higher than  that in  globular
clusters,  (ii) very close, for the  lowest luminosity dwarfs, to that
observed in the MW field and open  clusters, (iii) the Carina specific
$F^{\rm   BSS}_{\rm HB}$ frequency     probably sets a threshold   for
star-forming galaxies, and most interestingly, (iv)  shows a hint of a
$F^{\rm BSS}_{\rm HB}-{ M_{V}}$ anti-correlation.
This almost precludes the occurrence of  collisional binaries in dwarf
galaxies, and one may  conclude that  genuine  BSS sequences  in dwarf
galaxies are mainly made of primordial binaries.

Lastly, it  is  interesting  to note   how  the BSS frequency  in  the
low-luminosity  dwarfs      and open  clusters    (${\rm  \log}(N_{\rm
BSS}/N_{\rm  HB})\sim0.3-0.4$) is very close  to  that derived for the
Galactic halo (${\rm \log}(N_{\rm BSS}/N_{\rm HB})\sim0.6$) by Preston
\& Sneden.
The latter   value however  has  been derived  relying on  a composite
sample of only  62 blue metal-poor  stars that are  (i) distributed at
different line  of  sights;  (ii)  at  different distances; and   most
importantly,  (iii)  for which  no observational   BSS-HB star-by-star
correspondence can be established.
Thus, allowing for all these uncertainties in the field BSS frequency,
it is safe to conclude that the  observed open clusters-dwarf galaxies
BSS frequency sets  a realistic, and observational  upper limit to the
primordial BSS frequency in stellar systems.

\begin{acknowledgements}
We thank Belokurov V., Willman  B., Carrera R., Monelli M. and
Aparicio  A. for providing us their photometric catalogs.

\end{acknowledgements}

\bibliographystyle{aa}

\end{document}